\documentclass[11pt]{article}
\def\one{1\hskip -.37em 1}     
\def\proj{E\hskip -.69em I}     
\begin{document}
\begin{titlepage}
\begin{flushright}
%hep-th/yymmxxx\\
August 2003\\
Revised April 2004
\end{flushright}
\begin{centering}
 
{\ }\vspace{1cm}
 
{\Large\bf Revisiting the Fradkin-Vilkovisky Theorem}\\

\vspace{1.5cm}

Jan Govaerts$^{a,c)}$ and Frederik G. Scholtz$^{b,c)}$

\vspace{1.3cm}

$^{a)}${\em Institute of Nuclear Physics}\\
{\em Department of Physics, Catholic University of Louvain}\\
{\em 2, Chemin du Cyclotron, B-1348 Louvain-la-Neuve, Belgium}\\
{\tt jan.govaerts@fynu.ucl.ac.be}

\vspace{1.0cm}

$^{b)}${\em Institute of Theoretical Physics}\\
{\em Department of Physics, University of Stellenbosch}\\
{\em Stellenbosch 7600, South Africa}\\
{\tt fgs@sun.ac.za}

\vspace{1.0cm}

$^{c)}${\em Stellenbosch Institute for Advanced Study (STIAS)}\\
{\em Private Bag X1, 7602 Stellenbosch, South Africa}\\
{\tt http://www.stias.ac.za}

\vspace{1.0cm}

\begin{abstract}

\noindent The status of the usual statement of the Fradkin-Vilkovisky 
theorem, claiming complete independence of the Batalin-Fradkin-Vilkovisky 
path integral on the gauge fixing ``fermion" even within a nonperturbative
context, is critically reassessed. Basic, but subtle reasons why this 
statement cannot apply as such in a nonperturbative quantisation of
gauge invariant theories are clearly identified. A criterion
for admissibility within a general class of gauge fixing conditions is provided
for a large ensemble of simple gauge invariant systems. This criterion confirms
the conclusions of previous counter-examples to the usual statement of the
Fradkin-Vilkovisky theorem.

\end{abstract}

\vspace{10pt}

\end{centering} 

\vspace{25pt}

\end{titlepage}

\setcounter{footnote}{0}

\section{\bf Introduction}
\label{Sect1}

Among available approaches towards the quantisation of locally gauge invariant
systems, the general BRST quantisation methods are certainly the most popular
and widely used. Within the BRST-BFV Hamiltonian setting\cite{FV}, one result 
stands out as being most relevant, namely the so-called 
Fradkin-Vilkovisky (FV) theorem according to which, in its statement 
as usually given\cite{FV,MH}, the BRST invariant BFV path integral (BFV-PI) 
representation of transition amplitudes is totally independent
of the choice of gauge fixing conditions, the latter thus being made 
to one's best convenience. However in this form, such a claim has been 
disputed on different grounds\cite{JG1,JG2,JG,FS,AR}, while general classes 
of explicit counter-examples have been presented\cite{JG1,JG2,JG,JG3} within 
simple gauge invariant systems.

Indeed, all these examples agree with the following facts, which are to 
be considered as defining the actual content of the FV theorem\cite{JG1,JG}.
Given the gauge invariance properties built into the formalism, the BFV-PI is,
by construction, manifestly BRST and gauge invariant. Consequently, whatever
the choice of gauge fixing conditions being implemented, the BFV-PI always 
reduces to some integral over the space of gauge orbits of the original 
gauge invariant system. In particular, any two sets of gauge fixing 
conditions which are gauge transforms of one another lead to the same final 
result for the BFV-PI. Nevertheless, which ``covering" (an integration 
domain with some measure) of the space of gauge orbits is thereby selected, 
depends directly on the gauge equivalence class of gauge fixing conditions 
to which the specific choice of gauge fixing functions belongs. In other 
words, the BFV-PI depends on the choice of gauge fixing conditions only 
through the gauge equivalence class to which these conditions belong. 
Nonetheless, the BFV-PI cannot be totally independent of the choice of 
gauge fixing conditions. Gauge invariance of the BFV-PI is a necessary
condition, but it is not a sufficient one for a choice of gauge fixing 
conditions to be admissible. Indeed, an admissible gauge fixing is one 
whose gauge equivalence class defines a single covering of the space of 
gauge orbits, namely such that each of these orbits are included with 
equal nonvanishing weight in the final integration. Nonadmissibility, 
namely a Gribov problem\cite{Gribov}, arises whenever either some orbits 
are counted with a smaller or larger weight than others (Gribov problem of 
Type I), or when some orbits are not included at all (Gribov problem
of Type II), or both\cite{JG}. Since the identification of a general criterion
to characterise admissibility of arbitrary gauge fixing conditions appears to
be difficult at least\cite{FS,AR}, this issue is best addressed on a 
case by case basis.

Notwithstanding the explicit examples confirming the more precise statement of
the FV theorem as just described, the arguments purporting to establish 
complete independence of the BFV-PI on the choice of gauge fixing seem to be 
so general and transparent, being based on the nilpotency of the BRST charge 
and BRST invariance of the external states for which the BFV-PI is computed, 
that the usual FV theorem statement is most often just simply taken for 
granted and to be perfectly undisputable. Confronted with this contradictory 
situation, it is justified to reconsider the status of the FV theorem and 
identify the subtle reasons why the formal arguments do not
apply as usually described. This is the purpose of the present note, at least
within a general class of simple constrained systems to be described in 
Sect.\ref{Sect2.1}.

One should point out in this context that there is no reason to
question the validity of the usual statement of the FV theorem within
the restricted context of ordinary perturbation theory for Yang-Mills
theories. Indeed, there exists explicit and independent proof of this
fact\cite{Voronov}. Furthermore, perturbation theory amounts to considering 
a set of gauge orbits in the immediate vicinity of the gauge orbit belonging to the trivial gauge configuration. However, Gribov problems and nonperturbative gauge fixing
issues involve the larger topological properties of the space of gauge 
orbits\cite{Singer}, and it is within this context that the relevance of 
the FV~theorem is addressed in the present note. There is no doubt that in 
the case of Yang-Mills theories, for example, such issues must play a vital 
role when it comes to the nonperturbative topological features of strongly
interacting nonlinear dynamics.

The outline of this note is follows. After having described in 
Sect.\ref{Sect2} the general class of gauge invariant systems to be
considered, including their quantisation within Dirac's approach which is
free of any gauge fixing procedure, Sect.3 addresses their BRST quantisation.
Based on the usual plane wave representation of the Lagrange multiplier
sector of the extended phase space within that context, the actual content
of the FV theorem is critically reassessed within a general class
of gauge fixing conditions, while subtle aspects explaining
why its usual statement fails to apply are pointed out.
Then in Sect.\ref{Sect4}, a regularisation procedure for the Lagrange 
multiplier sector is considered, which avoids the use of the 
non-normalisable plane wave states, by compactifying that degree of
freedom into a circle. A general admissibility criterion for the classes
of gauge fixing conditions considered is then identified, while further
subtle reasons explaining why the usual statement of the FV theorem fails
also in that context are again pointed out. No inconsistencies between 
the two considered approaches arise, confirming the actual and precise 
content of the Fradkin-Vilkovisky theorem as given above. Concluding 
remarks are presented in Sect.5.

\section{A Simple General Class of Models}
\label{Sect2}

\subsection{Classical formulation}
\label{Sect2.1}

Let us consider a system whose configuration space is spanned by a set of
bosonic coordinates $q^n$, with canonically conjugate momenta denoted
$p_n$, thus with the canonical brackets $\{q^n,p_m\}=\delta^n_m$. These phase
space degrees of freedom are subjected to a single first-class constraint
$\phi(q^n ,p_n)=0$, which defines a local gauge invariance for such a
system. Finally, dynamics is generated from a first-class Hamiltonian 
$H(q^n,p_n)$, which we shall assume to have a vanishing bracket with 
the constraint $\phi$, $\{H,\phi\}=0$. Given that large classes of examples 
fall within such a description, the latter condition is only a mild 
restriction, which is made to ease some of the explicit evaluations 
to be discussed hereafter.

A well known\cite{JG} system meeting all the above requirements is that of the
relativistic scalar particle, in which case the first-class constraint
$\phi$ defines both the local generator for world-line
diffeomorphisms as well as the mass-shell condition for the particle
energy-momentum. Other examples in which the first-class constraint
is the generator of a local internal U(1) gauge invariance may easily be
imagined, such as those discussed Refs.\cite{JG3,GK}. In the latter reference
for instance, one has a collection of degrees of freedom $q^a_i(t)$ 
($a=1,2; i=1,2,\cdots, d$) with Lagrange function
\begin{equation}
L=\frac{1}{2}\left[\dot{q}^a_i-\lambda\epsilon^{ab}q^b_i\right]^2-
\frac{1}{2}\omega^2q^a_iq^a_i\ \ \ ,\ \ \ \epsilon^{ab}=-\epsilon^{ba}\ .
\label{eq:oscillator}
\end{equation}
This system may be interpreted as that of $d$ spherical harmonic oscillators
in a plane subjected to the constraint that their total angular momentum 
vanishes at all times,
\begin{equation}
\phi=\epsilon^{ab}p^a_iq^b_i=0\ \ \ \ ,\ \ \ 
p^a_i=\dot{q}^a_i-\lambda\epsilon^{ab}q^b_i\ .
\end{equation}
The U(1) gauge invariance of the system is that of arbitrary time-dependent
rotations in the plane acting identically on all oscillators, with $\lambda(t)$ 
being both the associated Lagrange multiplier and U(1) gauge degree of 
freedom (the time component of the gauge ``field").

Returning to the general setting, all the above characteristics
may be condensed into one single information, namely the
first-order Hamiltonian action principle over phase space expressed as
\begin{equation}
S[q^n,p_n;\lambda]=\int\,dt\,\left[\dot{q}^np_n-H-\lambda\,\phi\right],
\end{equation}
where $\lambda(t)$ is an arbitrary Lagrange multiplier associated to the 
first-class constraint $\phi(q^n,p_n)=0$. The Hamiltonian 
equations of motion are generated from the total Hamiltonian 
$H_T=H+\lambda\phi$, in which the Lagrange multiplier parametrises 
the freedom associated to small gauge transformations throughout 
time evolution of the system. These small gauge transformations 
are generated by the first-class constraint $\phi(q^n,p_n)$.

Indeed, in their infinitesimal form, small gauge transformations are generated
by the first-class constraint as
\begin{equation}
\delta_\epsilon q^n=\epsilon\{q^n,\phi\}\ \ \ ,\ \ \
\delta_\epsilon p_n=\epsilon\{p_n,\phi\}\ \ \ ,\ \ \ 
\delta_\epsilon \lambda=\dot{\epsilon},
\end{equation}
$\epsilon(t)$ being an arbirary function of time (the above action then changes 
only by a total time derivative). Related to this simple character of gauge 
transformations, it is readily established\cite{JG1,JG} that, given a choice of 
boundary conditions (b.c.) for which the coordinates $q^n(t)$ are specified at 
the boundary of some time interval $[t_i,t_f]$ ($t_i<t_f$), which then
also requires that the gauge transformation function obeys the b.c. 
$\epsilon(t_{i,f})=0$, the space of gauge orbits is in one-to-one 
correspondence with Teichm\"uller space, {\sl i.e.\/} the space of gauge 
orbits for the Lagrange multiplier $\lambda(t)$. In the present instance, 
this Teichm\"uller space reduces to the real line spanned by the gauge 
invariant modular or Teichm\"uller parameter
\begin{equation}
\gamma=\int_{t_i}^{t_f}\,dt\,\lambda(t).
\end{equation}
Consequently, any admissible gauge fixing of the system is thus to induce 
a covering of this modular space in which each of all the possible real 
values for $\gamma$ is accounted for with an equal weight. Indeed, any
real value for $\gamma$ characterises in a unique manner a possible gauge 
orbit of the system, while on the other hand any configuration of the system
belongs to a given gauge orbit. Thus, in order to account for all possible 
physically distinct gauge invariant configurations of the system, all 
possible values for the single coordinate parameter $\gamma$ on modular space
must be accounted for in any given admissible gauge fixing procedure
(absence of a Gribov problem of Type I), while at the same time none of
these orbits may be included with a weight that differs from that of any of
the other gauge orbits (absence of a Gribov problem of Type II).
An admissible gauge fixing procedure must induce a covering of modular space
which includes all real values for $\gamma$ with a $\gamma$-independent
integration measure over modular space.

\subsection{Quantum formulation}
\label{Sect2.2}

As the above notation makes already implicit, in order to avoid
any ambiguity in the forthcoming discussion, the configuration space manifold
is assumed to be of countable discrete dimension, if not simply finite. 
Furthermore at the quantum level, we shall also assume that the associated 
Hilbert space of quantum states itself is spanned by a discrete basis of states. 
Depending on the system, this may require to compactify configuration space,
such as for instance into a torus topology, or introduce some further 
interaction potential, such as a harmonic well, it being understood that
such regularisation procedures may be removed at the very end of the analysis.
In this manner, typical problems associated to plane wave representations of
the Heisenberg algebra, $[\hat{q}^n,\hat{p}_m]=i\hbar\delta^n_m$ with
$\hat{q}^{n\dagger}=\hat{q}^n$ and ${\hat{p}_n}^\dagger=\hat{p}_n$, are avoided 
from the outset. As a matter of fact, a torus regularisation procedure will 
be applied to the Lagrange multiplier sector when considering BRST quantisation
at some later stage of our discussion.

Furthermore, for ease of expression hereafter, we shall assume to be working in
a basis of Hilbert space which diagonalises the first-class constraint 
operator $\hat{\phi}$,
\begin{equation}
\hat{\phi}|k\rangle=\phi_k|k\rangle,
\end{equation}
with in particular the integers $k_0$ denoting the subset of these states
which is associated to a vanishing eigenvalue for the constraint 
with an unspecified degeneracy,
\begin{equation}
\hat{\phi}|k_0\rangle=0\ \ \ ,\ \ \ \phi_{k_0}=0.
\end{equation}
The latter states $|k_0\rangle$ for all the possible values $k_0$ thus
define a basis for the subspace of gauge invariant or physical states, which 
are to be annihilated by the constraint.

The examples mentioned in (\ref{eq:oscillator}) provide explicit illustrations
of such a general setting. The spectra of both the Hamiltonian and constraint
eigenstates are discrete, with specific degeneracies for each class, including
the physical sector of gauge invariant states. In the case of the relativistic
scalar particle, the same situation arises provided one introduces a 
regulating harmonic potential term quadratic in the spacetime coordinates 
in order to render the spectrum discrete. Even for a system as simple as 
a topological particle on a circle, for which the Lagrange function is given by 
$L=N\dot{q}$ where $N$ is some normalisation factor that needs to take
on a quantised value at the quantum level, the momentum constraint
operator, $\hat{\phi}=\hat{p}-N$, then also possesses a discrete spectrum,
and thus falls within the general setting of systems addressed in our
discussion (in this case, the first-class Hamiltonian vanishes identically,
while the gauge invariance associated to the first-class constraint is that
of arbitrary coordinate redefinitions of the degree of freedom $q(t)$).

Given an arbitrary choice for the Lagrange multiplier $\lambda(t)$, and since 
it is also assumed that quantisation preserves the gauge invariance property 
of the first-class Hamiltonian $\hat{H}$ (namely that even at the quantum level
we still have the vanishing commutator $[\hat{H},\hat{\phi}]=0$, which also
implies that the time-ordered exponential of the total Hamiltonian,
$\hat{H}_T(t)=\hat{H}+\lambda(t)\hat{\phi}$, coincides with its ordinary 
exponential), time evolution of the quantum system is generated by the operator
\begin{equation}
\hat{U}(t_f,t_i)=
e^{-\frac{i}{\hbar}\int_{t_i}^{t_f}dt\left[\hat{H}+\lambda(t)\hat{\phi}\right]},
\end{equation}
which propagates both gauge variant and invariant states. Propagation of 
physical states only is achieved by introducing the physical projection 
operator\cite{Klaud1} $\proj\,$, obtained essentially by integrating over 
the gauge group of all finite small gauge transformations
$e^{-i/\hbar\gamma\hat{\phi}}$, which in the present case may be expressed as
\begin{equation}
\proj=\lim_{\gamma_0\rightarrow\infty}\int_{-\gamma_0}^{\gamma_0}
\frac{d\gamma}{2\gamma_0}\, e^{-\frac{i}{\hbar}\gamma\hat{\phi}}=
\sum_{k_0}|k_0\rangle\langle k_0|\ \ \ ,\ \ \
\proj\,^2=\proj\ \ \ ,\ \ \ \proj\,^\dagger=\proj\,.
\end{equation}
Consequently, the physical evolution operator is given by
$\hat{U}_{\rm phys}(t_f,t_i)=\hat{U}(t_f,t_i)\proj=
\proj\,\hat{U}(t_f,t_i)\proj\,$, for which all matrix elements in the basis 
$|k\rangle$ vanish, except on the physical subspace spanned by the states
$|k_0\rangle$,
\begin{equation}
\begin{array}{r l}
&\cdot\ {\rm If}\ k_i\ne k_0\ {\rm or}\ k_f\ne k_0:\
\langle k_f|\hat{U}_{\rm phys}(t_f,t_i)|k_i\rangle=0;\\
& \\
&\cdot\ {\rm If}\ k_i=k_{0,i}\ {\rm and}\ k_f=k_{0,f}:\
\langle k_{0,f}|\hat{U}_{\rm phys}(t_f,t_i)|k_{0,i}\rangle=
\langle k_{0,f}|e^{-\frac{i}{\hbar}\Delta t\hat{H}}|k_{0,i}\rangle\ \ \ ,\ \ \
\Delta t=t_f-t_i.
\end{array}
\label{eq:prop1}
\end{equation}
The latter are thus the matrix elements that the BFV-PI must reproduce 
from BRST quantisation given any admissible gauge fixing choice.

Note that one may also write,
\begin{equation}\hat{U}_{\rm phys}(t_f,t_i)=
\lim_{\gamma_0\rightarrow\infty}\int_{-\gamma_0}^{\gamma_0}
\frac{d\gamma}{2\gamma_0}\,e^{-\frac{i}{\hbar}
\left[\Delta t\hat{H}+\gamma\hat{\phi}\right]}=
e^{-\frac{i}{\hbar}\Delta t\hat{H}}\,
\lim_{\gamma_0\rightarrow\infty}\int_{-\gamma_0}^{\gamma_0}
\frac{d\gamma}{2\gamma_0}\,e^{-\frac{i}{\hbar}\gamma\hat{\phi}},
\label{eq:prop2}
\end{equation}
which clearly reproduces the above matrix elements, and makes it explicit 
that one indeed has performed an admissible integration over the modular 
space of the system parametrised by $-\infty<\gamma<+\infty$ with a 
uniform integration measure\cite{JG4}, precisely a covering of modular
space which is characteristic of an admissible gauge fixing choice.

\section{BFV-BRST Formulation}
\label{Sect3}

\subsection{BFV extended phase space}
\label{Sect3.1}

Within the BFV approach\cite{FV,MH,JG}, phase space is first extended by 
introducing a momentum $\pi(t)$ canonically conjugate to the Lagrange multiplier
$\lambda(t)$, $\{\lambda(t),\pi(t)\}=1$. Consequently, one then has the
set of first class constraints $G_a=(G_1,G_2)=(\pi,\phi)=0$, $a=1,2$,
such that $\{H,G_a\}=0$. To compensate for these additional dynamical
degrees of freedom, a further system of pairs of Grassmann odd canonically
conjugate ghost degrees of freedom, $\eta^a(t)$ and ${\cal P}_a(t)$ with
$\eta^{a\dagger}=\eta^a$, ${\cal P}^\dagger_a=-{\cal P}_a$ and
$\{\eta^a,{\cal P}_b\}=-\delta^a_b$, is introduced. By convention,
$\eta^a$ (resp. ${\cal P}_a$) are of ghost number $+1$ (resp. $-1$).
The ghost number is given by $Q_g={\cal P}_a\eta^a$.

Within this setting, small local gauge transformations are traded for
global BRST transformations, generated by the BRST charge $Q_B$, which
in the present situation is simply given by
\begin{equation}
Q_B=\eta^aG_a=\eta^1\pi+\eta^2\phi,
\end{equation}
a Grassmann odd quantity, real under complex conjugation and of ghost 
number $(+1)$, characterised by its nilpotency property, $\{Q_B,Q_B\}=0$.

BRST invariant dynamics on this extended phase space is generated by the
general BRST invariant Hamiltonian
\begin{equation}
H_{\rm eff}=H-\{\Psi,Q_B\},
\end{equation}
$\Psi$ being an {\sl a priori\/} arbitrary Grassmann odd function of extended
phase space, pure imaginary under complex conjugation and of ghost number $(-1)$,
known as the ``gauge fixing fermion" as this is indeed the role it takes
within this formalism.

In order to obtain a BRST invariant dynamics, the equations of motion generated
from $H_{\rm eff}$ must be supplemented with BRST invariant boundary conditions.
Considering BRST transformations,
\begin{equation}
\delta_B q^n=\{q^n,Q_B\}=\eta^2\{q^n,\phi\}\ \ ,\ \
\delta_B\lambda=\eta^1\ \ ,\ \ 
\delta_B\eta^1=0\ \ ,\ \ 
\delta_B\eta^2=0,
\end{equation}
\begin{equation}
\delta_B p_n=\{p_n,Q_B\}=\eta^2\{p_n,\phi\}\ \ ,\ \ 
\delta_B\pi=0\ \ ,\ \ 
\delta_B{\cal P}_1=-\pi\ \ ,\ \ 
\delta_B{\cal P}_2=-\phi,
\end{equation}
it appears that a choice of b.c. which is universally BRST invariant
is such that
\begin{equation}
\pi(t_{i,f})=0\ \ \ ,\ \ \ 
{\cal P}_1(t_{i,f})=0\ \ \ ,\ \ \ 
\eta^2(t_{i,f})=0,
\label{eq:BRSTbc}
\end{equation}
while the b.c. in the original ``matter" sector $(q^n,p_n)$ are
those already mentioned in the discussion of Sect.\ref{Sect2.1}.
Note that since $Q_B(t_{i,f})=0$ as well as
$Q_g(t_{i,f})=0$, while $\dot{Q}_B=\{Q_B,H_{\rm eff}\}=0$
and $\dot{Q}_g=\{Q_g,H_{\rm eff}\}=0$ on account of the BRST invariance 
and vanishing ghost number of $H_{\rm eff}$, these b.c. imply that any 
solution is indeed BRST invariant and of vanishing ghost number, $Q_B(t)=0$ 
and $Q_g(t)=0$. These are precisely the b.c. that are imposed in the 
construction of the BFV-PI for the BRST invariant quantised system.

Obviously, a condition which the gauge fixing function $\Psi$ must meet is that,
given the above b.c., the set of solutions to the equations of motion generated
by the corresponding Hamiltonian $H_{\rm eff}$ coincides exactly with the set
of solutions obtained in the initial formulation of Sect.\ref{Sect2.1}. This
requirement restricts already on the classical level the classes of gauge 
fixing functions $\Psi$ that may be considered. Even the classical BRST 
invariant dynamics is not entirely independent of the choice of 
$\Psi$\cite{JG}, a point we shall not pursue further here (having already 
been discussed to some extent in Ref.\cite{JG} through detailed examples), 
but which indicates that it cannot be so either at the quantum level.

A general class of functions that is to be used explicitly hereafter
within the quantised system is of the form
\begin{equation}
\Psi={\cal P}_1F(\lambda)+\beta{\cal P}_2\lambda,
\label{eq:Psi}
\end{equation}
$F(\lambda)$ being an arbitrary real function and $\beta$ an arbitrary real 
parameter. It may readily be established that associated to this choice,
the classical Hamiltonian equation of motion for $\lambda(t)$ amounts to
the gauge fixing condition
\begin{equation}
\frac{d\lambda(t)}{dt}=F(\lambda)\ .
\label{eq:lambda}
\end{equation}
In terms of some integration constant $\lambda_0$, the solution 
$\lambda(t;\lambda_0)$ defines a value $\gamma(\lambda_0)$
for the Teichm\"uller parameter. Given a choice for $F(\lambda)$, as the value
$\lambda_0$ varies over its domain of definition, $\gamma(\lambda_0)$ varies
over a certain domain in modular space with a specific oriented covering
or measure over that domain. It is only when the entire set of real values
for the Teichm\"uller parameter $\gamma$ is obtained with a $\gamma$-independent
integration measure that the function $F(\lambda)$, namely $\Psi$, defines
an admissible gauge fixing choice.

For instance, the case $F(\lambda)=0$ is readily seen to meet this 
admissibility requirement, and to define a choice of gauge fixing which is 
known to be admissible for the considered class of 
systems\cite{MH,JG1,JG,Teit}. Indeed, the equation of motion for $\lambda$ 
then simply reads $\dot{\lambda}=0$, showing that all values of the 
Teichm\"uller parameter $\gamma$ are obtained with a single multiplicity 
when integrating over the free integration constant $\lambda_0=\lambda(t_0)$ 
at some time value $t=t_0$, the b.c. in this sector being $\pi(t_{i,f})=0$. 
One of the purposes of the present note is to identify, at the quantum level, 
a general criterion for the admissibility of the class of gauge fixing 
functions in (\ref{eq:Psi}).

\subsection{BRST quantisation}
\label{Sect3.2}

Quantisation of the BFV formulation amounts to constructing a linear 
representation space for the (anti)commutation relations,
\begin{equation}
\left[\hat{q}^n,\hat{p}_m\right]=i\hbar\delta^n_m\ \ \ ,\ \ \
\left[\hat{\lambda},\hat{\pi}\right]=i\hbar\ \ \ ,\ \ \
\left\{\hat{c}^a,\hat{b}_b\right\}=\delta^a_b,
\end{equation}
with $\hat{\eta}^a=\hat{c}^a$ and $\hat{\cal P}_a=-i\hbar\hat{b}_a$, and 
equiped with an hermitean inner product $\langle{\ }\cdot|\cdot{\ }\rangle$ 
such that all these operators are self-adjoint. Note that $\hat{c}^{a2}=0$ 
and $\hat{b}^2_a=0$.

Quantisation of the ``matter" sector $(q^n,p_n)$ has already been dealt with
in Sect.\ref{Sect2.2}, for which we shall use the same notations and choice 
of basis. An abstract representation space for the ghost sector $(c^a,b_a)$ 
is constructed as follows\cite{JG}: that sector of Hilbert space is spanned 
by a basis with $2^2=4$ vectors denoted $|\pm\pm\rangle$ (the first entry 
refering to the sector $a=1$ and the second to the sector $a=2$; this
convention also applies to the bra-states $\langle\pm\pm|$), on which 
the ghost operators act as follows,
\begin{equation}
\begin{array}{l l l l l}
&\hat{c}^1|--\rangle=|+-\rangle\ \ ,\ \ 
&\hat{c}^1|+-\rangle=0\ \ ,\ \ 
&\hat{c}^1|-+\rangle=|++\rangle\ \ ,\ \ 
&\hat{c}^1|++\rangle=0,\\
& & & & \\
&\hat{c}^2|--\rangle=|-+\rangle\ \ ,\ \ 
&\hat{c}^2|+-\rangle=-|++\rangle\ \ ,\ \ 
&\hat{c}^2|-+\rangle=0\ \ ,\ \ 
&\hat{c}^2|++\rangle=0,\\
& & & & \\
&\hat{b}_1|--\rangle=0\ \ ,\ \ 
&\hat{b}_1|+-\rangle=|--\rangle\ \ ,\ \ 
&\hat{b}_1|-+\rangle=0\ \ ,\ \ 
&\hat{b}_1|++\rangle=|-+\rangle,\\
& & & & \\
&\hat{b}_2|--\rangle=0\ \ ,\ \ 
&\hat{b}_2|+-\rangle=0\ \ ,\ \ 
&\hat{b}_2|-+\rangle=|--\rangle\ \ ,\ \ 
&\hat{b}_2|++\rangle=-|+-\rangle\,.
\end{array}
\end{equation}
Their only nonvanishing inner products are
\begin{equation}
\langle --|++\rangle=
-\langle +-|-+\rangle=
\langle -+|+-\rangle=
-\langle ++|--\rangle,
\end{equation}
with any of these numbers pure imaginary, such as
for instance $\langle --|++\rangle=\pm i$. Finally, the
normal ordered quantum ghost number operator is defined as
\begin{equation}
\hat{Q}_g=\frac{1}{2}\left[\hat{c}^a\hat{b}_a-\hat{b}_a\hat{c}^a\right]
\ \ \ ,\ \ \ 
\hat{Q}^\dagger_g=-\hat{Q}_g.
\end{equation}
Consequently, one has the following ghost number values for these states,
\begin{equation}
\hat{Q}_g|--\rangle=(-1)|--\rangle\ \ \ ,\ \ \
\hat{Q}_g|+-\rangle=0\ \ \ ,\ \ \
\hat{Q}_g|-+\rangle=0\ \ \ ,\ \ \
\hat{Q}_g|++\rangle=(+1)|++\rangle.
\end{equation}

Even though at some later stage in our discussion we shall perform a
circle compactification of the Lagrange multiplier degree of freedom,
let us at this point consider the usual plane wave representation of the
Heisenberg algebra in the $(\hat{\lambda},\hat{\pi})$ Lagrange multiplier
sector. Eigenstates of these operators are thus defined by
\begin{equation}
\hat{\lambda}|\lambda\rangle = \lambda|\lambda\rangle\ \ \ ,\ \ \ 
-\infty<\lambda<+\infty\ \ \ ;\ \ \ 
\hat{\pi}|\pi\rangle = \pi|\pi\rangle\ \ \ ,\ \ \ 
-\infty<\pi<+\infty\ , 
\end{equation}
with the normalisation choices,
\begin{equation}
\langle\lambda|\lambda'\rangle=\delta(\lambda-\lambda')\ \ \ ,\ \ \ 
\langle\pi|\pi'\rangle=\delta(\pi-\pi')\ \ \ ;\ \ \ 
\one=\int_{-\infty}^\infty d\lambda|\lambda\rangle\langle\lambda|=
\int_{-\infty}^\infty d\pi|\pi\rangle\langle\pi|\ .
\end{equation}
Consequently, one has the wave function representations of these operators
acting on any state $|\psi\rangle$,
\begin{equation}
\langle\lambda|\hat{\lambda}|\psi\rangle=\lambda\,\langle\lambda|\psi\rangle
\ \ \ ,\ \ \ 
\langle\lambda|\hat{\pi}|\psi\rangle=-i\hbar\frac{\partial}{\partial\lambda}\,
\langle\lambda|\psi\rangle
\ \ \ ;\ \ \ 
\langle\pi|\hat{\lambda}|\psi\rangle=i\hbar\frac{\partial}{\partial\pi}
\,\langle\pi|\psi\rangle
\ \ \ ,\ \ \ 
\langle\pi|\hat{\pi}|\psi\rangle=\pi\,\langle\pi|\psi\rangle\ ,
\end{equation}
with the matrix elements for the change of basis,
\begin{equation}
\langle\lambda|\pi\rangle=\frac{1}{\sqrt{2\pi\hbar}}\,
e^{\frac{i}{\hbar}\lambda\,\pi}\ \ \ ,\ \ \ 
\langle\pi|\lambda\rangle=\frac{1}{\sqrt{2\pi\hbar}}\,
e^{-\frac{i}{\hbar}\lambda\,\pi}\ .
\end{equation}

The quantum BRST charge is given by
\begin{equation}
\hat{Q}_B=\hat{c}^1\hat{\pi}+\hat{c}^2\hat{\phi}\ \ \ ,\ \ \ 
\hat{Q}^2_B=0\ \ \ ,\ \ \ 
\hat{Q}^\dagger_B=\hat{Q}_B.
\end{equation}
Furthermore, time evolution of the quantised system is generated by the BRST
invariant Hamiltonian operator
\begin{equation}
\hat{H}_{\rm eff}=\hat{H}+\frac{i}{\hbar}\left\{\hat{\Psi},\hat{Q}_B\right\},
\end{equation}
leading to the BRST invariant evolution operator
\begin{equation}
\hat{U}_{\rm eff}(t_f,t_i)=e^{-\frac{i}{\hbar}\Delta t\hat{H}_{\rm eff}}.
\end{equation}
For the class of gauge fixing functions (\ref{eq:Psi}), an explicit 
evaluation finds
\begin{equation}
\hat{H}_{\rm eff}=\hat{H}+\beta\hat{\lambda}\hat{\phi}+
\frac{1}{2}\left[F(\hat{\lambda})\hat{\pi}+\hat{\pi}F(\hat{\lambda})\right]
+\frac{1}{2}\left[F(\hat{\lambda})\hat{\pi}-\hat{\pi}F(\hat{\lambda})\right]
\left[\hat{b}_1\hat{c}^1-\hat{c}^1\hat{b}_1\right]
+i\hbar\beta\hat{b}_2\hat{c}^1,
\label{eq:Heff}
\end{equation}
this operator being expressed in such a way as to make manifest its
hermiticity property, $\hat{H}^\dagger_{\rm eff}=\hat{H}_{\rm eff}$.

Classically within the extended formulation, physical states need to meet the
constraints $\pi(t)=0$ and $\phi(t)=0$, which implies that for the BRST 
quantised system, the BRST invariance conditions characterising physical 
states must lead to the eigenvalues $\phi_k=0$ and $\pi=0$, namely $k=k_0$ 
and $\pi=0$. This is achieved by considering the cohomology of the 
BRST charge, {\sl i.e.\/} by considering the states which are BRST invariant 
but are defined modulo a BRST transformation,
\begin{equation}
|\psi\rangle = |\psi_{\rm phys}\rangle\,+\,\hat{Q}_B|\varphi\rangle\ \ \ ,\ \ \ 
\hat{Q}_B|\psi\rangle=0.
\end{equation}
It may be shown that the general solution to this equation is given by
\begin{equation}
\begin{array}{r c l}
|\psi_{\rm phys}\rangle&=&
\sum_{k_0}\psi_{k_0;--}(\pi=0)\,|k_0;\pi=0;--\rangle\\
& & \\
& &+\sum_{k_0}\left\{\psi_{k_0;+-}(\pi=0)\,|k_0;\pi=0;+-\rangle\ +\
\psi_{k_0;-+}(\pi=0)\,|k_0;\pi=0;-+\rangle\right\}\\
 & & \\
& & +\sum_{k_0}\psi_{k_0;++}(\pi=0)\,|k_0;\pi=0;++\rangle,
\end{array}
\label{eq:BRSexact}
\end{equation}
while the state $|\varphi\rangle$ may be constructed from the remaining
components of the BRST invariant state $|\psi\rangle$ expanded in the basis 
$|k;\pi;\pm\pm\rangle$,
$|\psi\rangle=\sum_{k;\pm\pm}\int_{-\infty}^\infty\,d\pi\,
\psi_{n;\pm\pm}(\pi)\,|n;\pi;\pm\pm\rangle$.
Consequently, both the BRST cohomology classes at the smallest and largest
ghost numbers, $\hat{Q}_g=-1$ and $\hat{Q}_g=+1$, are in one-to-one 
correspondence with the physical states $|k_0\rangle$ in Dirac's 
quantisation (or $|k_0;\pi=0\rangle$ when the Lagrange multiplier sector 
is included), while the BRST cohomology class at zero ghost number, 
$\hat{Q}_g=0$, includes two copies of the Dirac physical states, associated 
to each of the ghost states $|+-\rangle$ and $|-+\rangle$. Physical states 
are usually defined to correspond to the BRST cohomology class at zero 
ghost number\cite{MH}.

The matrix elements of the BRST invariant evolution operator
$\hat{U}_{\rm eff}(t_f,t_i)$ between states of ghost number $(-1)$ all vanish
identically, on account of the vanishing ghost number of $\hat{H}_{\rm eff}$
and the vanishing inner product $\langle --|--\rangle=0$,
\begin{equation}
\langle k_f;\pi_f;--|\hat{U}_{\rm eff}(t_f,t_i)|k_i;\pi_i;--\rangle=0,
\label{eq:physical0}
\end{equation}
irrespective of the choice of gauge fixing function $\Psi$, and whether
the external states of ghost number $(-1)$ are BRST invariant or not.

However, these are not the matrix elements of $\hat{U}_{\rm eff}(t_f,t_i)$ that
ought to correspond to those in (\ref{eq:prop1}) and (\ref{eq:prop2})
which describe in Dirac's quantisation the propagation of physical states only.
Indeed, the latter may be obtained only for external states which are 
BRST invariant and of vanishing ghost number, in direct correspondence with 
the choice of such b.c. in (\ref{eq:BRSTbc}). Equivalently, given the action 
of the ghost and BRST operators, such states are spanned by the set 
$|k;\pi=0;-+\rangle$, so that we now have to address the explicit evaluation 
of the matrix elements
\begin{equation}
\langle k_f;\pi_f=0;-+|\hat{U}_{\rm eff}(t_f,t_i)|k_i;\pi_i=0;-+\rangle.
\label{eq:ME1}
\end{equation}
By construction, these matrix elements are clearly BRST and thus gauge 
invariant, and include those of the BRST cohomology class at zero ghost 
number associated to one of the two sets of states corresponding to Dirac's 
physical states. Nevertheless, these matrix elements are not totally 
independent of the choice of gauge fixing function $\Psi$, as shall 
now be established. 

\subsection{The BFV-BRST invariant propagator}
\label{Sect3.3}

Given the choice of gauge fixing function in (\ref{eq:Psi}) and the expression 
for the associated Hamiltonian $\hat{H}_{\rm eff}$ in (\ref{eq:Heff}), it is 
clear that (\ref{eq:ME1}) factorizes into two contributions, whether the
conditions $\pi_f=0=\pi_i$ required for BRST invariance of the
external states are enforced or not,
\begin{equation}
\langle k_f;\pi_f;-+|\hat{U}_{\rm eff}(t_f,t_i)|k_i;\pi_i;-+\rangle=
\langle k_f|e^{-\frac{i}{\hbar}\Delta t\hat{H}}|k_i\rangle\,\times\,
{\cal N}(\pi_f,\pi_i;\phi_{k_i}),
\end{equation}
with the factor ${\cal N}(\pi_f,\pi_i;\phi_{k_i})$ given 
by\footnote{Note that in this expression one could as well replace the 
value $\phi_{k_i}$ by $\phi_{k_f}$.}
\begin{equation}
{\cal N}(\pi_f,\pi_i;\phi_{k_i})=\langle \pi_f;-+|e^{-\frac{i}{\hbar}\Delta t
\left[\beta\phi_{k_i}\hat{\lambda}+
\frac{1}{2}\left(F(\hat{\lambda})\hat{\pi}+\hat{\pi}F(\hat{\lambda})\right)+
\frac{1}{2}\left(F(\hat{\lambda})\hat{\pi}-\hat{\pi}F(\hat{\lambda})\right)
\left(\hat{b}_1\hat{c}^1-\hat{c}^1\hat{b}_1\right)+
i\hbar\beta\hat{b}_2\hat{c}^1\right]}
|\pi_i;-+\rangle .
\end{equation}
Of course, one is particularly interested in the value for 
${\cal N}(\pi_f=0,\pi_i=0;\phi_{k_i})$ as function of the first-class 
constraint spectral value $\phi_{k_i}$ or $\phi_{k_f}$.

As a warm-up, let us first restrict to the choice $F(\lambda)=0$, known
to be admissible. On basis of the above explicit expression for
${\cal N}(\pi_f,\pi_i;\phi_{k_i})$, it is clear that in this case, 
one has the further factorisation
\begin{equation}
{\cal N}(\pi_f,\pi_i;\phi_{k_i})=
\langle -+|e^{\beta\Delta t\hat{b}_2\hat{c}^1}|-+\rangle\,
\langle\pi_f|e^{-\frac{i}{\hbar}\beta\Delta t\phi_{k_i}\hat{\lambda}}|
\pi_i\rangle,
\end{equation}
whose value readily reduces to
\begin{equation}
{\cal N}(\pi_f,\pi_i;\phi_{k_i})=-\beta\Delta t\langle -+|+-\rangle\,
\delta(\pi_i-\pi_f-\beta\Delta t\phi_{k_i})\ .
\label{eq:Nvalue1}
\end{equation}
Restricting then to the BRST invariant external states, one has finally (the
condition $\Delta t>0$ is implicit),
\begin{equation}
\langle k_f;\pi_f=0;-+|\hat{U}_{\rm eff}(t_f,t_i)|k_i;\pi_i=0;-+\rangle =
-{\rm sgn}(\beta)\,\langle -+|+-\rangle\,\delta(\phi_{k_i})
\langle k_f|e^{-\frac{i}{\hbar}\Delta t\hat{H}}|k_i\rangle\ .
\label{eq:Nvalue0}
\end{equation}
Hence indeed, all these matrix elements vanish identically, unless both
external states are physical, namely $k_i=k_{0,i}$ and $k_f=k_{0,f}$ or
$\phi_{k_i}=0=\phi_{k_f}$. However, when the external states are physical,
these matrix elements are singular, on account of the $\delta$-function
$\delta(\phi_{k_i})$. Clearly, this is a direct consequence of the plane wave
representation of the Heisenberg algebra in the Lagrange multiplier sector
$(\hat{\lambda},\hat{\pi})$ of the BFV extended phase space. Nevertheless, 
up to the singular normalisation factor 
$(-{\rm sgn}(\beta)\langle -+|+-\rangle\delta(\phi_{k_i}))$, the BFV-BRST
invariant matrix elements reproduce correctly the result in (\ref{eq:prop1})
for the propagator of physical states within Dirac's quantisation approach.
On the other hand, note also that this distribution-valued normalisation
factor is not entirely independent of the choice of function $\Psi$ even
when $F(\lambda)=0$, since it depends on the sign of the arbitrary parameter
$\beta$. In spite of that dependency, an admissible gauge fixing is
achieved since all of modular space is indeed recovered with a 
$\gamma$-independent integration measure.

Turning now to an arbitrary choice of function $F(\lambda)$, the explicit
and exact evaluation of ${\cal N}(\pi_f,\pi_i;\phi_{k_i})$ may proceed 
through its discretized path integral representation. Applying the approach 
detailed in Ref.\cite{JG}, one then establishes the general and exact result,
\begin{equation}
{\cal N}(\pi_f=0,\pi_i=0;\phi_{k_i})=-\beta\langle -+|+-\rangle\,
\int_{{\cal D}[F]}\frac{d\gamma}{2\pi\hbar}\,
e^{-\frac{i}{\hbar}\beta\gamma\phi_{k_i}},
\end{equation}
where the domain of integration ${\cal D}[F]$ in modular space
is identified as follows. Given a choice for the function $F(\lambda)$ and 
thus the gauge fixing condition in (\ref{eq:lambda}), the solution 
$\lambda(t;\lambda_f)$ is obtained as function of the integration
constant $\lambda_f=\lambda(t_f;\lambda_f)$ at the final value of the
time interval $[t_i,t_f]$, thereby leading to a specific value
$\gamma(\lambda_f)$ for the Teichm\"uller parameter. As the integration 
constant $\lambda_f$ varies over its entire domain of definition, 
$-\infty<\lambda_f<+\infty$, the modular parameter $\gamma(\lambda_f)$ then
defines a certain domain ${\cal D}[F]$ in modular space, including the
orientation induced by the sign of $d\gamma(\lambda_f)/d\lambda_f$ in
the case of multicoverings. This is how the choice of gauge fixing 
function $F(\lambda)$ determines a specific covering of modular space, 
namely a specific domain ${\cal D}[F]$ in $\gamma$ together with a specific
integration measure. This is precisely the manner\cite{JG1,JG2,JG} in which 
the gauge invariant BFV-PI is dependent on the choice of gauge 
fixing fermion function $\Psi$, in contradiction with the usual 
statement\cite{MH} of the Fradkin-Vilkovisky theorem.

An admissible choice of gauge fixing is thus associated to
${\cal D}[F]$ being the entire real line, in which case,
\begin{equation}
{\cal N}(\pi_f=0,\pi_i=0;\phi_{k_i})=
-\beta\langle -+|+-\rangle\,\delta(\beta\phi_{k_i})=
-{\rm sgn}(\beta)\,\langle -+|+-\rangle\,\delta(\phi_{k_i})\ ,
\end{equation}
thus reproducing indeed the result (\ref{eq:Nvalue0}) established 
for $F(\lambda)=0$. A more general class of admissible gauge choices 
is given by
\begin{equation}
F(\lambda)=a+b\lambda,
\label{eq:ce1}
\end{equation}
$a$ and $b$ being constant parameters. On the other hand, choices such as
\begin{equation}
F(\lambda)=a+b\lambda+c\lambda^2\ (c\ne 0)\ \ \ ,\ \ \ 
F(\lambda)=a\lambda^3\ (a> 0)\ \ \ ,\ \ \ 
F(\lambda)=e^{-a\lambda}\ (a>0)\ ,
\label{eq:ce2}
\end{equation}
all define gauge fixing choices which are not admissible\cite{JG1,JG2,JG,JG3}.
For instance when $F(\lambda)=a\lambda^3$, the modular domain ${\cal D}[F]$
is finite and given by the interval $[-\sqrt{2\Delta t/a},\sqrt{2\Delta t/a}]$
in modular space. The BFV-PI is thus indeed dependent on
the choice of gauge fixing function, albeit in a gauge invariant manner.
Nonetheless, in the limit that $a\rightarrow 0$, an admissible covering
of modular space is recovered, associated to the choice $F(\lambda)=0$.

\subsection{Deconstructing the Fradkin-Vilkovisky theorem}
\label{Sect3.4}

An argument often invoked\cite{MH} in support of complete independence of the
BFV-PI on the choice of gauge fixing fermion is based on
the observation that for BRST invariant external states $|\psi_1\rangle$
and $|\psi_2\rangle$ such that $\hat{Q}_B|\psi_i\rangle=0$ ($i=1,2$), the 
matrix elements of the operator $\{\hat{\Psi},\hat{Q}_B\}$ vanish identically,
\begin{equation}
\langle\psi_1|\{\hat{\Psi},\hat{Q}_B\}|\psi_2\rangle=
\langle\psi_1|\left[\hat{\Psi}\hat{Q}_B+\hat{Q}_B\hat{\Psi}\right]
|\psi_2\rangle=0,
\label{eq:argument1}
\end{equation}
where the last equality follows by considering the separate action of
the BRST operator $\hat{Q}_B$ on the external states adjacent to it.
Indeed, given nilpotency of the BRST charge, $\hat{Q}^2_B=0$, this argument
should also extend to similar matrix elements of the evolution operator
$\hat{U}_{\rm eff}(t_f,t_i)$ which includes the contribution,
\begin{equation}
e^{-\frac{i}{\hbar}\Delta t\frac{i}{\hbar}\left\{\hat{\Psi},\hat{Q}_B\right\}}=
\one\ +\ \frac{\Delta t}{\hbar^2}
\left[\hat{\Psi}\hat{Q}_B+\hat{Q}_B\hat{\Psi}\right]\ +\
\frac{1}{2!}\left(\frac{\Delta t}{\hbar^2}\right)^2
\left[\hat{\Psi}\hat{Q}_B\hat{\Psi}\hat{Q}_B+
\hat{Q}_B\hat{\Psi}\hat{Q}_B\hat{\Psi}\right]\ +\ \cdots\ .
\label{eq:argument2}
\end{equation}

In the case of the factor ${\cal N}(\pi_f,\pi_i;\phi_{k_i})$, this argument 
would appear to imply that one should have, for the states of interest,
\begin{equation}
\langle \pi_f=0;-+|e^{-\frac{i}{\hbar}\Delta t\frac{i}{\hbar}
\left\{\hat{\Psi},\hat{Q}_B\right\}}|\pi_i=0;-+\rangle=
\langle \pi_f=0;-+|\pi_i=0;-+\rangle=\delta(0)\cdot 0\ ,
\end{equation}
given the facts that $\langle \pi_f|\pi_i\rangle=\delta(\pi_f-\pi_i)$ and
$\langle -+|-+\rangle=0$. Even though this expression is ill-defined,
it appears to be totally independent of the choice of gauge fixing fermion
$\Psi$, in sharp contrast with its previous evaluations.

The singular character of this result follows once again from the plane wave
representation of the Lagrange multiplier sector $(\hat{\lambda},\hat{\pi})$.
Consequently, matrix elements are generally distribution-valued, and
cannot simply be evaluated at specific values of their arguments. Rather, they
should be convolved with test functions, or else evaluated first
for arbitrary values of their arguments\cite{JG}. Hence the above argument 
certainly cannot be claimed to be standing on a sound basis, and needs to be 
reconsidered carefully for the explicit evaluation of 
${\cal N}(\pi_f,\pi_i;\phi_{k_i})$ given a specific value $\phi_{k_i}$ for the 
constraint eigenvalue but as yet unspecified values for $\pi_f$ and $\pi_i$.

In order to remain faithful to the spirit of the above argument, the calculation
needs to be performed in the form as given in (\ref{eq:argument2}), namely
not by first computing the result of the anticommutator 
$\{\hat{\Psi},\hat{Q}_B\}$ and only then compute its matrix elements---in 
effect, this is the procedure used to reach the results of 
Sect.\ref{Sect3.3}---, but rather by having the operators act from left 
to right onto the external state $|\psi_2\rangle$ for the first term 
inside the square brackets at each order in $\Delta t/\hbar^2$ in 
(\ref{eq:argument2}), and from right to left onto the state 
$\langle\psi_1|$ for the second term. This calculation is straightforward 
for the specific admissible choice $F(\lambda)=0$, in which case one obtains,
\begin{equation}
{\cal N}(\pi_f,\pi_i;\phi_{k_i})=i\hbar\beta(\pi_i-\pi_f)\langle -+|+-\rangle
\sum_{k=1}^\infty\frac{1}{k!}\left(\frac{\Delta t}{\hbar^2}\right)^k\,
\left(-i\hbar\beta\phi_{k_i}\right)^{k-1}\,
\langle\pi_f|\hat{\lambda}^k|\pi_i\rangle\ .
\end{equation}
It would appear that indeed, this expression vanishes whenever one
considers BRST invariant external states for which $\pi_f=0=\pi_i$.
However, this is not the case, since the factor which is multiplied
by $(\pi_i-\pi_f)$ is itself singular for the values $\pi_f=0=\pi_i$, 
being distribution-valued. Indeed, the above sum may also be expressed as
\begin{equation}
\begin{array}{r l}
{\cal N}(\pi_f,\pi_i;\phi_{k_i})=
&-\frac{1}{\phi_{k_i}}\,\langle -+|+-\rangle\,(\pi_i-\pi_f)\,
\langle \pi_f|\sum_{k=1}^\infty\frac{1}{k!}
\left(\frac{\Delta t}{\hbar^2}\right)^k
\left(-i\hbar\beta\phi_{k_i}\right)^k\hat{\lambda}^k|\pi_i\rangle\\
 & \\
=&-\frac{1}{\phi_{k_i}}\langle -+|+-\rangle\,(\pi_i-\pi_f)\,
\langle\pi_f|
\left[e^{-\frac{i}{\hbar}\Delta t\beta\phi_{k_i}\hat{\lambda}}-1\right]
|\pi_i\rangle\\
 & \\
=&-\frac{1}{\phi_{k_i}}\langle -+|+-\rangle\,(\pi_i-\pi_f)\,
\left[\delta(\pi_i-\pi_f-\beta\Delta t\phi_{k_i})-\delta(\pi_i-\pi_f)\right]\\
 & \\
=&-\beta\Delta t\langle -+|+-\rangle\,
\delta(\pi_i-\pi_f-\beta\Delta t\phi_{k_i}),
\end{array}
\label{eq:Nvalue2}
\end{equation}
a result that coincides with (\ref{eq:Nvalue1}). Nevertheless, the details of
the calculation in the above series of relations make manifest the fact that
had one set from the outset the values $\pi_f=0=\pi_i$, an identically 
vanishing result would have been obtained, rather than the correct but
distribution-valued one, ${\cal N}(\pi_f=0,\pi_i=0;\phi_{k_i})=
-\beta\Delta t\langle -+|+-\rangle\delta(\beta\Delta t\phi_{k_i})$, which
does vanish unless when precisely $\phi_{k_i}=0$. On the other hand, if 
from the outset one considers a value $\phi_{k_i}=0$, one finds through 
the above analysis,
\begin{equation}
\begin{array}{r l}
{\cal N}(\pi_f,\pi_i;\phi_{k_i})=&
i\hbar\beta(\pi_i-\pi_f)\langle -+|+-\rangle\frac{\Delta t}{\hbar^2}\,
\langle\pi_f|\hat{\lambda}|\pi_i\rangle\\
 & \\
=&i\hbar\beta(\pi_i-\pi_f)\langle -+|+-\rangle\frac{\Delta t}{\hbar^2}\,
\left(-i\hbar\frac{\partial}{\partial\pi_i}\delta(\pi_i-\pi_f)\right)\\
 & \\
=& -\beta\Delta t\langle -+|+-\rangle\,\delta(\pi_i-\pi_f)\ ,
\end{array}
\end{equation}
once again in agreement with the general results in (\ref{eq:Nvalue1})
and (\ref{eq:Nvalue2}). However, performing such a calculation with
$\pi_f=0=\pi_i$ from the outset leads back to an identically vanishing
result, missing once again the correct distribution-valued result,
${\cal N}(\pi_f=0,\pi_i=0;\phi_{k_i})=
-\beta\Delta t\langle -+|+-\rangle\delta(\beta\Delta t\phi_{k_i})$.

In conclusion, these considerations establish that the argument
based on (\ref{eq:argument1}) or (\ref{eq:argument2}), purportedly
a confirmation that the BFV-PI is necessarily totally
independent of the gauge fixing fermion $\Psi$, is not warranted.
Being distribution-valued quantities, the relevant matrix elements
have to be convolved with test functions, or equivalently, first
be evaluated for whatever external states, and only at the end
restricted to the BRST invariant ones. In particular, setting from the
outset the values $\pi_f=0=\pi_i$ is ill-fated, indeed even leads
to ill-defined quantities such as $0\cdot\delta(0)$. Nevertheless, when
properly computed, the end result is perfectly consistent with that
established in the previous section in a totally independent manner.
And in the latter approach, a general expression for 
${\cal N}(\pi_f,\pi_i;\phi_{k_i})$ is even amenable to an exact
evaluation for whatever choice of function $F(\lambda)$, through a path
integral representation of the matrix elements of relevance. This exact
result displays explicitly the full extent to which, in a manner totally 
consistent with the built-in gauge invariance properties of the BFV-PI, 
the gauge fixed BFV-BRST path integral is indeed dependent on the
choice of gauge fixing fermion $\Psi$\cite{JG1,JG2,JG}, namely only through
the gauge equivalence class to which that gauge fixing choice belongs,
such a gauge equivalence class being characterised by a specific covering
of modular space. Being gauge invariant, the BFV-PI necessarily
reduces to an integral over modular space, irrespective of the gauge fixing
choice. Nevertheless, which domain and integration measure over modular
space are thereby induced are function of the choice of gauge fixing conditions.
The BFV-PI is not totally independent of the choice of gauge fixing 
fermion $\Psi$.

\section{The Admissibility Criterion}
\label{Sect4}

As manifest from previous expressions, the plane wave representation
of the Heisenberg algebra in the Lagrange multiplier sector 
$(\hat{\lambda},\hat{\pi})$ leads to distribution-valued results for
specific BFV-BRST matrix elements. Consequently, it is sometimes 
claimed\cite{MH2} that this very fact calls into question the relevance of the
counter-examples to the usual statement of the FV theorem available in the
literature and described in the previous sections, while a proper
handling of the ensuing singularities would show that these counter-examples
are actually ill-fated, and that indeed, the BFV-PI ought to be totally 
independent of the choice of gauge fixing fermion $\Psi$.

In order to avoid having to deal with non-normalisable plane wave states,
let us now regularise the Lagrange multiplier sector by compactifying the 
degree of freedom $\lambda$ onto a circle of circumference $2L$ such that 
$-L\le\lambda<L$, it being understood that any quantity of interest has to 
be evaluated in the decompactification limit $L\rightarrow\infty$. Furthermore, 
the representation of the Heisenberg algebra 
$[\hat{\lambda},\hat{\pi}]=i\hbar$, 
which is to be used on this space with the nontrivial mapping class group 
$\pi_1(S_1)=Z\!\!\!Z$, is that of vanishing U(1) holonomy\footnote{A 
representation of nonvanishing U(1) holonomy may also be used,
provided the BRST charge is given a quantum correction linear in 
the holonomy in order to preserve its nilpotency, thereby preserving all 
our conclusions.}\cite{JG5}. Consequently, this sector of Hilbert space is 
now spanned by a discrete set of $\hat{\pi}$-eigenstates for all integer 
values $m$,
\begin{equation}
\hat{\pi}|m\rangle=\pi_m|m\rangle\ \ \ ,\ \ \ 
\pi_m=\frac{\pi\hbar}{L}m\ \ \ ,\ \ \ 
\langle m|m'\rangle=\delta_{m,m'}\ \ \ ,\ \ \
\one=\sum_{m}|m\rangle\langle m|.
\end{equation}
The configuration space wave functions are
\begin{equation}
\langle\lambda|m\rangle=\frac{1}{\sqrt{2L}}e^{i\frac{\pi m}{L}\lambda}
\ \ \ ,\ \ \ 
\langle m|\lambda\rangle=\frac{1}{\sqrt{2L}}e^{-i\frac{\pi m}{L}\lambda},
\label{eq:wavec}
\end{equation}
$|\lambda\rangle$ being the configuration space basis such that
\begin{equation}
\hat{\lambda}|\lambda\rangle=\lambda|\lambda\rangle\ \ ,\ \ -L\le\lambda<L
\ \ ,\ \ 
\langle\lambda|\lambda'\rangle=\delta_{2L}(\lambda-\lambda')\ \ ,\ \ 
\one=\int_{-L}^L d\lambda|\lambda\rangle\langle\lambda|.
\label{eq:operatorc}
\end{equation}
Given an arbitrary state $|\psi\rangle$ and its configuration
space wave function $\psi(\lambda)=\langle\lambda|\psi\rangle$ which must
be single-valued on the circle, one has
\begin{equation}
\langle\lambda|\hat{\lambda}|\psi\rangle=\lambda\,\psi(\lambda)\ \ \ ,\ \ \ 
\langle\lambda|\hat{\pi}|\psi\rangle=-i\hbar\frac{d}{d\lambda}\,\psi(\lambda).
\end{equation}
States in this sector are thus characterised by the normalisibility
condition $\int_{-L}^L d\lambda|\psi(\lambda)|^2<\infty$. In the above 
relations, $\delta_{2L}(\lambda-\lambda')$ stands for the $\delta$-function 
on the circle of circumference $2L$,
\begin{equation}
\delta_{2L}(\lambda-\lambda')=
\frac{1}{2L}\sum_{m}e^{i\frac{\pi m}{L}(\lambda-\lambda')}.
\end{equation}

Given such a discretization of the Lagrange multiplier sector 
$(\hat{\lambda},\hat{\pi})$, let us now address again the different points
raised previously concerning the FV theorem.

BRST cohomology classes remain characterised in the same way as
previously. The general solution to the BRST invariance condition
$\hat{Q}_B|\psi\rangle=0$, namely $|\psi\rangle=|\psi_{\rm phys}\rangle\,+\,
\hat{Q}_B|\varphi\rangle$, is given by
\begin{equation}
\begin{array}{r c l}
|\psi_{\rm phys}\rangle&=&
\sum_{k_0}\psi_{k_0;m=0;--}\,|k_0;m=0;--\rangle\\
& & \\
& &+\sum_{k_0}\left\{\psi_{k_0;m=0;+-}\,|k_0;m=0;+-\rangle\ +\
\psi_{k_0;m=0;-+}\,|k_0;m=0;-+\rangle\right\}\\
 & & \\
& & +\sum_{k_0}\psi_{k_0;m=0;++}\,|k_0;m=0;++\rangle,
\end{array}
\end{equation}
while the state $|\varphi\rangle$ may be constructed from the remaining
components of the BRST invariant state $|\psi\rangle$ expanded in 
the basis $|k;m;\pm\pm\rangle$, 
$|\psi\rangle=\sum_{k;m;\pm\pm}\psi_{k;m;\pm\pm}\,|k;m;\pm\pm\rangle$.
Consequently, both the BRST cohomology classes at the smallest and largest
ghost numbers, $\hat{Q}_g=-1$ and $\hat{Q}_g=+1$, are in one-to-one 
correspondence with the physical states $|k_0\rangle$ in Dirac's quantisation 
(or $|k_0;m=0\rangle$ when the Lagrange multiplier sector is included), while
the BRST cohomology class at zero ghost number, $\hat{Q}_g=0$, includes two
copies of the Dirac physical states, associated to each of the ghost
states $|+-\rangle$ and $|-+\rangle$. Physical states are usually defined
to correspond to the BRST cohomology class at zero ghost number\cite{MH}.

The matrix elements of the BRST invariant evolution operator
$\hat{U}_{\rm eff}(t_f,t_i)$ between states of ghost number $(-1)$ all vanish
identically, on account of the vanishing ghost number of $\hat{H}_{\rm eff}$
and the vanishing inner product $\langle --|--\rangle=0$,
\begin{equation}
\langle k_f;m_f;--|\hat{U}_{\rm eff}(t_f,t_i)|k_i;m_i;--\rangle=0,
\label{eq:physical}
\end{equation}
irrespective of the choice of gauge fixing function $\Psi$, and whether
the external states of ghost number $(-1)$ are BRST invariant or not.

However, these are not the matrix elements of $\hat{U}_{\rm eff}(t_f,t_i)$ that
ought to correspond to those in (\ref{eq:prop1}) and (\ref{eq:prop2})
which describe in Dirac's quantisation the propagation of physical states only.
Indeed, the latter may be obtained only for external states which are 
BRST invariant and of vanishing ghost number, in direct correspondence 
with the choice of such b.c. in (\ref{eq:BRSTbc}). Equivalently, given 
the action of the ghost and BRST operators, such states are spanned by 
the set $|k;m=0;-+\rangle$, so that we now have to address the explicit 
evaluation of the matrix elements
\begin{equation}
\langle k_f;m_f=0;-+|\hat{U}_{\rm eff}(t_f,t_i)|k_i;m_i=0;-+\rangle,
\label{eq:ME}
\end{equation}
the discretized analogues of the matrix elements in (\ref{eq:ME1}).
As before these matrix elements are, by construction, BRST and thus gauge 
invariant, and include those of the BRST cohomology class at zero ghost 
number associated to one of the two sets of states corresponding to 
Dirac's physical states. Nevertheless, they are not totally independent
of the choice of gauge fixing function $\Psi$, as shall now be 
established once again.

\subsection{Evaluation of the BRST invariant matrix elements}
\label{Sect4.1}

In order to evaluate the matrix elements (\ref{eq:ME}), rather than using
a path integral approach, the operator representation of the quantised
system shall be considered. Given the choice of gauge fixing function 
in (\ref{eq:Psi}) and the expression for the associated Hamiltonian 
$\hat{H}_{\rm eff}$ in (\ref{eq:Heff}), it is clear that (\ref{eq:ME}) 
as well as its extension for whatever values for $m_f$ and $m_i$
factorizes as
\begin{equation}
\langle k_f;m_f;-+|\hat{U}_{\rm eff}(t_f,t_i)|k_i;m_i;-+\rangle=
\langle k_f|e^{-\frac{i}{\hbar}\Delta t\hat{H}}|k_i\rangle\,\times\,
{\cal N}_L(m_f,m_i;\phi_{k_i}),
\end{equation}
with the factor ${\cal N}_L(m_f,m_i;\phi_{k_i})$ given 
by\footnote{Note that in this expression
one could as well replace the value $\phi_{k_i}$ by $\phi_{k_f}$.}
\begin{equation}
\begin{array}{r l}
&{\cal N}_L(m_f,m_i;\phi_{k_i})=\\
 & \\
&=\langle m_f;-+|e^{-\frac{i}{\hbar}\Delta t
\left[\beta\phi_{k_i}\hat{\lambda}+
\frac{1}{2}\left(F(\hat{\lambda})\hat{\pi}+\hat{\pi}F(\hat{\lambda})\right)+
\frac{1}{2}\left(F(\hat{\lambda})\hat{\pi}-\hat{\pi}F(\hat{\lambda})\right)
\left(\hat{b}_1\hat{c}^1-\hat{c}^1\hat{b}_1\right)+
i\hbar\beta\hat{b}_2\hat{c}^1\right]}|m_i;-+\rangle .
\end{array}
\end{equation}
The evaluation of the ghost contribution to this factor, through a direct 
expansion of the exponential operator and a resolution of the ensuing 
recurrence relations, implies a further factorization
\begin{equation}
\begin{array}{r l}
&{\cal N}_L(m_f,m_i;\phi_{k_i})=-i\hbar\beta\langle -+|+-\rangle\times\\
 & \\
&\times\langle m_f|
\sum_{n=0}^\infty\frac{1}{(n+1)!}\left(-\frac{i}{\hbar}\Delta t\right)^{n+1}
\sum_{k=0}^n
\left(\beta\phi_{k_i}\hat{\lambda}+\hat{\pi}F(\hat{\lambda})\right)^k
\left(\beta\phi_{n_i}\hat{\lambda}+F(\hat{\lambda})\hat{\pi}\right)^{n-k}
|m_i\rangle .
\end{array}
\end{equation}

Consider then the quantities
\begin{equation}
\left[\beta\phi_{k_i}\hat{\lambda}+F(\hat{\lambda})\hat{\pi}\right]^n
|m=0\rangle=G_n(\hat{\lambda})|m=0\rangle\ \ \ ,\ \ \ n=0,1,2,\cdots,
\end{equation}
where the functions $G_n(\lambda)$ are defined by their relation to
the l.h.s. operator acting on the state $|m=0\rangle$.
These functions obey the recurrence relations
\begin{equation}
G_{n+1}(\lambda)=\beta\phi_{k_i}\lambda G_n(\lambda)
-i\hbar F(\lambda)\frac{dG_n(\lambda)}{d\lambda}\ \ \ ,\ \ \ 
G_0(\lambda)=1.
\end{equation}
Introducing the variable $u$ such that
\begin{equation}
\frac{d\lambda(u)}{du}=F\left(\lambda(u)\right),
\end{equation}
given some initial value $\lambda_0=\lambda(u_0)$, the functions $G_n(\lambda)$
are solved by
\begin{equation}
G_n(\lambda)=e^{-\frac{i}{\hbar}\beta\phi_{k_i}\int_{u_0}^u dv\lambda(v)}
\left(-i\hbar\frac{d}{du}\right)^n\,
e^{\frac{i}{\hbar}\beta\phi_{k_i}\int_{u_0}^u dv\lambda(v)}.
\end{equation}
Using the representation
\begin{equation}
\left[\beta\phi_{k_i}\hat{\lambda}+F(\hat{\lambda})\hat{\pi}\right]^n
|m=0\rangle=\int_{-L}^L\frac{d\lambda}{\sqrt{2L}}|\lambda\rangle\,G_n(\lambda),
\end{equation}
it thus follows that one may write
\begin{equation}
{\cal N}_L(m_f=0,m_i=0;\phi_{k_i})=
-i\hbar\beta\langle -+|+-\rangle\,\int_{-L}^L\frac{d\lambda}{2L}
\sum_{n=0}^\infty\frac{1}{(n+1)!}\left(-\frac{i}{\hbar}\Delta t\right)^{n+1}
\sum_{k=0}^n G^*_k(\lambda)G_{n-k}(\lambda).
\label{eq:result}
\end{equation}

It is of interest to first consider the choice $F(\lambda)=0$, which is known
to define an admissible gauge fixing.  One then has 
$G_n(\lambda)=(\beta\phi_{k_i}\lambda)^n$, leading to the following values,
\begin{equation}
\begin{array}{l}
\cdot\ {\rm If}\ \phi_{k_i}=0:\ 
{\cal N}_L(m_f=0,m_i=0;\phi_{k_i})=-\beta\Delta t\langle -+|+-\rangle;\\
 \\
\cdot\ {\rm If}\ \phi_{k_i}\ne 0:\
{\cal N}_L(m_f=0,m_i=0;\phi_{k_i})=-\beta\Delta t\langle -+|+-\rangle\times
\frac{\sin\left(\beta\Delta t L\phi_{k_i}/\hbar\right)}
{\left(\beta\Delta t L\phi_{k_i}/\hbar\right)}.
\end{array}
\label{eq:MEL}
\end{equation}
Consequently, in the limit $L\rightarrow\infty$, the matrix elements 
(\ref{eq:ME}) are given by,
\begin{equation}
\begin{array}{r l}
&\cdot\ {\rm If}\ k_i\ne k_0\ {\rm or}\ k_f\ne k_0: \\
&\langle k_f;m_f=0;-+|\hat{U}_{\rm eff}(t_f,t_i)|k_i;m_i=0;-+\rangle=0;\\
 \\
&\cdot\ {\rm If}\ k_i=k_{0,i}\ {\rm and}\ k_f=k_{0,f}: \\
&\langle k_f;m_f=0;-+|\hat{U}_{\rm eff}(t_f,t_i)|k_i;m_i=0;-+\rangle=
\left[-\beta\Delta t\langle -+|+-\rangle\right]\
\langle k_{0,f}|e^{-\frac{i}{\hbar}\Delta t\hat{H}}|k_{0,i}\rangle.
\end{array}
\label{eq:MEadmissible}
\end{equation}
Hence indeed, up to a $\beta$-dependent normalisation, these matrix elements
reproduce those in (\ref{eq:prop1}) representing within Dirac's quantisation 
the propagation of physical states only. Given the representation in 
(\ref{eq:prop2}), one thus concludes that the choice $F(\lambda)=0$ defines 
an admissible gauge fixing.  

Let us now turn to the general case of an arbitrary function $F(\lambda)$. 
Given the result (\ref{eq:result}), it is clear that whenever 
$\phi_{k_i}=0$ and $\phi_{k_f}=0$, the matrix element (\ref{eq:ME}) reduces 
again to the same value as in (\ref{eq:MEL}) and
(\ref{eq:MEadmissible}). However, it is 
the decoupling of the unphysical states which may not be realised\cite{FS}, 
implying specific restrictions on the choice for $F(\lambda)$. In order 
to apply the limit $L\rightarrow\infty$ to these matrix elements, it is best 
to introduce a rescaled variable $\lambda=L\tilde{\lambda}$ with 
$-1\le\tilde{\lambda}<1$. Given the general expression (\ref{eq:result}), 
it should be clear that in order to reproduce the same results as in 
the admissible case $F(\lambda)=0$, the following limit
\begin{equation}
\lim_{L\rightarrow\infty}\frac{1}{L}F(L\tilde{\lambda})=
\tilde{F}(\tilde{\lambda})
\label{eq:criterion}
\end{equation}
is to define a finite function $\tilde{F}(\tilde{\lambda})$
of $\tilde{\lambda}$ for all values of $\tilde{\lambda}$. Whenever this 
criterion is met, the choice of gauge fixing function in (\ref{eq:Psi}) 
defines an admissible gauge fixing of the system, for which the BRST 
invariant matrix elements (\ref{eq:ME}) are given as in 
(\ref{eq:MEadmissible}), and do indeed
reproduce, up to some normalisation factor which is also function of the
parameter $\beta$, the correct time evolution of Dirac's physical states 
only. Note, however, that the resulting matrix elements in 
(\ref{eq:MEadmissible}) are nonetheless functions of the parameter $\beta$ 
appearing in such choices of admissible functions $\Psi$. Furthermore,
when the criterion (\ref{eq:criterion}) is not met, the associated choice of
gauge fixing is not admissible, since the BRST invariant matrix elements
(\ref{eq:ME}) then do not coincide with (\ref{eq:MEadmissible}), and thus 
cannot be expressed through a single integral covering of Teichm\"uller space
as in (\ref{eq:prop2}). In other words, the BFV-PI, which provides the phase
space path integral representation for the BRST invariant matrix elements
(\ref{eq:ME}), cannot be entirely independent of the choice of gauge fixing
``fermion" function $\Psi$, in contradiction with the FV theorem as 
usually stated.

The conclusion reached in (\ref{eq:criterion}) is also consistent with 
the explicit examples available in the literature and recalled in 
(\ref{eq:ce1}) and (\ref{eq:ce2}). Note that all these examples do indeed 
agree with the general criterion for admissibility established in 
(\ref{eq:criterion}).

\subsection{Deconstructing the Fradkin-Vilkovisky theorem in discretized form}
\label{Sect4.2}

Let us now address, within the discretized Lagrange multiplier sector,
the general argument claiming to confirm independence of the BFV-PI on
the choice of gauge fixing fermion, based on the expressions 
(\ref{eq:argument1}) and (\ref{eq:argument2}).

First consider again the states of ghost number $(-1)$, 
spanned by $|k;m;--\rangle$ for all values of $k$ and $m$. 
One readily finds
\begin{equation}
\langle k_1;m_1;--|\{\hat{\Psi},\hat{Q}_B\}|k_2;m_2;--\rangle=0,
\end{equation}
as it must since $\{\hat{\Psi},\hat{Q}_B\}$ is of zero ghost number while the
$(-1)$ ghost number sector is spanned only by $|--\rangle$ which is such that
$\langle --|--\rangle=0$. Note that this result also agrees with that
established in (\ref{eq:physical}), which applies for the same reasons. Thus
the conclusion in (\ref{eq:argument1}) is valid on the
BRST cohomology class at ghost number $(-1)$ on account of these simple and
general facts, totally independently of the choice for $\Psi$, and, for that 
matter, of the argument in (\ref{eq:argument1}) itself.

Let us now consider the BRST invariant states $|k;m=0;-+\rangle$ used in
the evaluation of the BFV-PI, and more generally the matrix elements of
the operator in (\ref{eq:argument2}), at a specific eigenvalue $\phi_{k_i}$
of the constraint $\hat{\phi}$, for the states $|m;-+\rangle$,
\begin{equation}
{\cal N}_L(m_f,m_i;\phi_{k_i})=
\langle m_f;-+|e^{-\frac{i}{\hbar}\Delta t\frac{i}{\hbar}
\left\{\hat{\Psi},\hat{Q}_B\right\}}|m_i;-+\rangle ,
\label{eq:Nvalue3}
\end{equation}
it being understood that the action of the operators on these external states
is evaluated along the same lines as in Sect.\ref{Sect3.4}. Hence, this
evaluation shall also be done for the specific choice $F(\lambda)=0$
known to be admissible and to lead to the results in (\ref{eq:MEL}) and
(\ref{eq:MEadmissible}).

The explicit expansion of the above matrix elements then reduces to the
following series of expressions, in perfect analogy with the calculation 
in (\ref{eq:Nvalue2}),
\begin{equation}
\begin{array}{r l}
{\cal N}_L(m_f,m_i;\phi_{k_i})
=&i\hbar\beta\frac{\pi\hbar}{L}(m_i-m_f)\langle -+|+-\rangle\,
\sum_{k=1}^\infty\frac{1}{k!}\left(\frac{\Delta t}{\hbar^2}\right)^k
\left(-i\hbar\beta\phi_{k_i}\right)^{k-1}
\langle m_f|\hat{\lambda}^k|m_i\rangle\\
 & \\
=&-\frac{1}{\phi_{k_i}}\frac{\pi\hbar}{L}(m_i-m_f)\langle -+|+-\rangle\,
\langle m_f|\left[e^{-\frac{i}{\hbar}\Delta t\beta\phi_{k_i}\hat{\lambda}}
-1\right]|m_i\rangle\\
 & \\
=&-\frac{1}{\phi_{k_i}}\frac{\pi\hbar}{L}(m_i-m_f)\langle -+|+-\rangle\,
\left\{\langle m_f|e^{-\frac{i}{\hbar}\Delta t\beta\phi_{k_i}\hat{\lambda}}
|m_i\rangle\,-\,\delta_{m_f,m_i}\right\}\\
 & \\
=&-\beta\Delta t\langle -+|+-\rangle\,\frac{1}{\beta\Delta t\phi_{k_i}}\,
\frac{\pi\hbar}{L}(m_i-m_f)\,\int_{-L}^L\frac{d\lambda}{2L}\,
e^{i\frac{\pi}{L}(m_i-m_f)\lambda}\,
e^{-\frac{i}{\hbar}\beta\Delta t\phi_{k_i}\lambda} .
\end{array}
\label{eq:evaluation2}
\end{equation}
Note that in this form, setting from the outset the values $m_f=0=m_i$
leads to a vanishing expression, as it did in the analysis of 
Sect.\ref{Sect3.4}. Furthermore, if from the outset we take the physical
values $\phi_{k_i}=0=\phi_{k_f}$, only the term with $k=1$ in the above sum
survives, leading to the following values,
\begin{equation}
\begin{array}{l}
\cdot\ {\rm If}\ m_f=m_i\ :\ 
{\cal N}_L(m_f,m_i;\phi_{k_i})=0;\\
 \\
\cdot\ {\rm If}\ m_f\ne m_i\ :\
{\cal N}_L(m_f,m_i;\phi_{k_i})=
\beta\Delta t\langle -+|+-\rangle\,\cos\pi(m_i-m_f).
\end{array}
\end{equation}
None of these results thus reproduce the correct ones in (\ref{eq:MEL}).
However, in the plane wave representations of Sect.\ref{Sect3.4} these
quantities being distribution-valued, a final integration by parts had
to be applied before recovering the correct result. Likewise in the
present discretized representation, the final evaluation of the above
expression finally leads to
\begin{equation}
\begin{array}{r l}
&{\cal N}_L(m_f,m_i;\phi_{k_i})=\\
 & \\
=&-\beta\Delta t\langle -+|+-\rangle\,
\left\{\int_{-L}^L\frac{d\lambda}{2L}\,
e^{i\frac{\pi}{L}(m_i-m_f)\lambda}\,
e^{-\frac{i}{\hbar}\beta\Delta t\phi_{k_i}\lambda}\ -\
e^{i\pi(m_i-m_f)}\,\frac{\sin\left(\beta\Delta t L\phi_{k_i}/\hbar\right)}
{\left(\beta\Delta t L \phi_{k_i}/\hbar\right)}\right\}.
\end{array}
\label{eq:evaluation3}
\end{equation}
Setting now $m_f=0=m_i$, still the value for
${\cal N}_L(m_f=0,m_i=0;\phi_{k_i})$ vanishes identically, irrespective
of whether the constraint eigenvalue $\phi_{k_i}$ is physical or not.
Nevertheless, by having compactified the degree of freedom $\lambda(t)$
onto a circle thus leading only to a discrete spectrum of quantum states
in the Lagrange multiplier sector $(\hat{\lambda},\hat{\pi})$,
we have avoided any use of distribution-valued matrix elements. 
Why, then, does the argument based on (\ref{eq:argument1}) and
(\ref{eq:argument2}) still not lead to the correct result?

The fact of the matter is that the adjoint action from the right
onto the external states $\langle m_f;-+|$ of the operators 
$(\hat{Q}_B\hat{\Psi}\hat{Q}_B\hat{\Psi}\cdots)$ in
(\ref{eq:argument2}) is not necessarily warranted when the operators 
$\hat{\lambda}$ and $\hat{\pi}$ appear in combination for the 
compactified regularisation. For example, consider the matrix elements
\begin{equation}
\langle m_f|\hat{\lambda}\hat{\pi}|m_i\rangle = \frac{\pi\hbar}{L}m_i\,
\langle m_f|\hat{\lambda}|m_i\rangle\ \ \ ,\ \ \ 
\langle m_f|\hat{\pi}\hat{\lambda}|m_i\rangle = \frac{\pi\hbar}{L}m_f\,
\langle m_f|\hat{\lambda}|m_i\rangle ,
\end{equation}
where in the second expression the adjoint action of the
operator $\hat{\pi}$ onto the state $\langle m_f|$ is used. However, one
must then conclude that
\begin{equation}
\langle m_f|\left[\hat{\lambda},\hat{\pi}\right]|m_i\rangle = 
\frac{\pi\hbar}{L}(m_i-m_f)\langle m_f|\hat{\lambda}|m_i\rangle\ ,
\end{equation}
in obvious contradiction with the Heisenberg algebra,
\begin{equation}
\langle m_f|\left[\hat{\lambda},\hat{\pi}\right]|m_i\rangle =
i\hbar\langle m_f|m_i\rangle = i\hbar\delta_{m_f,m_i}.
\end{equation}
In presence of the operator $\hat{\lambda}$, the adjoint action of $\hat{\pi}$
on bra-states should be avoided. Rather, one should evaluate the action
of all operators from the left onto ket-states and only at the very end
project the result onto the relevant bra-states. For instance,
\begin{equation}
\hat{\lambda}\hat{\pi}|m_i\rangle = \frac{\pi\hbar}{L}m_i\,
\hat{\lambda}|m_i\rangle\ \ \ ,\ \ \ 
\hat{\pi}\hat{\lambda}|m_i\rangle =-i\hbar|m_i\rangle\,+\,
\frac{\pi\hbar}{L}m_i\,\hat{\lambda}|m_i\rangle,
\label{eq:algebra2}
\end{equation}
so that
\begin{equation}
\langle m_f|\hat{\lambda}\hat{\pi}|m_i\rangle = \frac{\pi\hbar}{L}m_i\,
\langle m_f|\hat{\lambda}|m_i\rangle\ \ \ ,\ \ \ 
\langle m_f|\hat{\pi}\hat{\lambda}|m_i\rangle = -i\hbar\delta_{m_f,m_i}\,+\,
\frac{\pi\hbar}{L}m_i\, \langle m_i|\hat{\lambda}|m_i\rangle ,
\label{eq:relation2}
\end{equation}
in obvious agreement with the Heisenberg algebra 
$\left[\hat{\lambda},\hat{\pi}\right]=i\hbar$. The same conclusions
may be reached by considering the explicit wave function representations
of the Heisenberg algebra given in (\ref{eq:wavec}) and (\ref{eq:operatorc}) 
for the circle topology. In fact, the operator $\hat{\lambda}$ being
represented through multiplication by $\lambda$ of single-valued wave
functions $\langle\lambda|\psi\rangle$ on the circle for which the operator 
$\hat{\pi}=-i\hbar\partial/\partial\lambda$ is self-adjoint, leads to
wave functions that are no longer single-valued on the circle. 
In particular, the required integration by parts corresponding to 
the adjoint action of the derivative operator 
$\hat{\pi}=-i\hbar\partial/\partial\lambda$ induces a nonvanishing surface
term because of the lack of single-valuedness of the wave function
$\lambda\langle\lambda|\psi\rangle$, in direct correspondence with the
second relation in (\ref{eq:relation2}). 
In other words, even though both operators are well 
defined on the space of normalisable wave functions on the circle, 
the operator $\hat{\lambda}$ maps outside the domain of states for which 
the operator $\hat{\pi}$ is self-adjoint.

This is the thus the core reason why the evaluation of the matrix
element (\ref{eq:ME}) according to the argument in (\ref{eq:argument2})
in which the strings of operators 
$(\cdots\hat{\Psi}\hat{Q}_B\hat{\Psi}\hat{Q}_B)$ and
$(\hat{Q}_B\hat{\Psi}\hat{Q}_B\hat{\Psi}\cdots)$ act separately
from the left onto the ket-states $|m_i;-+\rangle$ and from the right
onto the bra-states $\langle m_f;-+|$, respectively, is unwarranted.
Indeed, even when $F(\lambda)=0$, precisely the combination
$\hat{\pi}\hat{\lambda}$ appears in the product $\hat{Q}_B\hat{\Psi}$
for which, as detailed above, the adjoint action of $\hat{\pi}$ 
from the right onto the bra-states is not justified unless the proper 
surface term contributions are accounted for as well
(whereas for the product $\hat{\Psi}\hat{Q}_B$ the relevant combination is 
$\hat{\lambda}\hat{\pi}$ which unambiguously acts from the left onto 
the ket-states).

Nevertheless, such ambiguities do not arise for the actual anticommutator
$\left\{\hat{\Psi},\hat{Q}_B\right\}$ when it is explicitly evaluated,
without keeping the two classes of terms separate as done in the
argument based on (\ref{eq:argument1}) and (\ref{eq:argument2}).
For example when $F(\lambda)=0$, the potentially troublesome term that is
then left over is simply
\begin{equation}
\left\{-i\hbar\beta\hat{b}_2\hat{\lambda},\hat{c}^1\hat{\pi}\right\}=
-i\hbar\beta\left[\hat{\lambda},\hat{\pi}\right]\hat{b}_2\hat{c}^1=
\hbar^2\beta\hat{b}_2\hat{c}^1,
\end{equation}
and is thus responsible for the transformation of the ghost ket-state 
$|-+\rangle$ into the state $|+-\rangle$ possessing a nonvanishing overlap 
with the ghost bra-state $\langle -+|$.

Applying this prescription for the evaluation of the matrix
elements in (\ref{eq:Nvalue3}), in fact one is brought back to the
approach used in Sect.\ref{Sect4.1}, thereby reproducing the general
results established in that context. For example when $F(\lambda)=0$,
a direct calculation along the lines of (\ref{eq:evaluation2}) readily finds
\begin{equation}
{\cal N}_L(m_f,m_i;\phi_{k_i})=-\beta\Delta t\langle -+|+-\rangle\,
\langle m_f|e^{-\frac{i}{\hbar}\beta\Delta t\phi_{k_i}\hat{\lambda}}|m_i\rangle,
\end{equation}
hence finally
\begin{equation}
\begin{array}{l}
\cdot\ {\rm If}\ \phi_{k_i}=0:\
{\cal N}_L(m_f,m_i;\phi_{k_i})=
-\beta\Delta t\,\langle -+|+-\rangle\,\delta_{m_f,m_i};\\
 \\
\cdot\ {\rm If}\ \phi_{k_i}\ne 0:\
{\cal N}_L(m_f,m_i;\phi_{k_i})=
-\beta\Delta t\,\langle -+|+-\rangle\,
\int_{-L}^L\frac{d\lambda}{2L}\,
e^{-\frac{i}{\hbar}\beta\Delta t\phi_{k_i}\lambda}\,
e^{i\frac{\pi}{L}(m_i-m_f)\lambda} ,
\end{array}
\end{equation}
a result to be compared to (\ref{eq:evaluation3}) in light of the remarks
in (\ref{eq:algebra2}) and (\ref{eq:relation2}). In particular, setting 
then $m_f=0=m_i$, exactly the same results as in (\ref{eq:MEL}) and 
(\ref{eq:MEadmissible}) in the $L\rightarrow\infty$ limit are thus recovered.

In conclusion, even though the compactification regularisation of the Lagrange
multiplier sector was introduced to circumvent the subtle issues explaining
why the argument, based on (\ref{eq:argument1}) and (\ref{eq:argument2})
and plane wave representations of the Heisenberg algebra and
claiming to confirm that the BFV-PI indeed ought to be totally independent
of the choice of gauge fixing fermion $\Psi$, is unwarranted, new subtleties
arise for a finite value of $L$ implying again that this argument does not
stand up to closer scrutiny. When properly analysed, the argument rather
confirms once again the results obtained by direct evaluation of the
relevant matrix elements. In particular, these matrix element 
corresponding to the BFV-PI, even though gauge invariant, are not
independent of the choice of gauge fixing procedure. The general
criterion for the admissibility of the class of gauge fixing fermions
defined in (\ref{eq:Psi}) is provided in (\ref{eq:criterion}).

\section{Conclusions}
\label{sect5}

Rather than gauge fixing the system through its Lagrange multiplier sector,
as is achieved through the choice made in (\ref{eq:Psi}), it is also possible
to contemplate gauge fixing in phase space through some condition of the
form $\chi(q^n,p_n)=0$, which, within the BFV-BRST formalism, is related to
the following choice of gauge fixing ``fermion",
\begin{equation}
\Psi=\rho{\cal P}_1\chi(q^n,p_n)+\beta{\cal P}_2\lambda,
\label{eq:Psi2}
\end{equation}
$\rho$ being an arbitrary real parameter. In the same manner as described
in this note for the class of gauge choices (\ref{eq:Psi}), it would be
of interest to identify a criterion that the function $\chi(q^n,p_n)$ should
meet in order that the associated gauge fixing be admissible. However,
this issue turns out to be quite involved, and we have not been able to
develop a general solution. In fact, in contradistinction to the class
of gauge fixings analysed in this note, the answer to this problem in
the case of the choices in (\ref{eq:Psi2}) would also depend on more
detailed properties of the first-class Hamiltonian $H$, the structure of
the original configuration space $q^n$, and how the local gauge transformations
generated by the first-class constraint $\phi$ act on that space.
In Ref.\cite{FS}, two specific models are considered for which the criterion
of admissibility in terms of the function $\chi(q^n,p_n)$ is indeed
different for each model.

Another simple model which was considered is defined by the seemingly 
trivial action
\begin{equation}
S[q]=\int dt N\dot{q},
\end{equation}
where the single degree of freedom $q(t)$ takes its values in a circle 
of radius $R$, while $N$ is some normalisation factor. The associated 
first-class constraint $\phi=p-N$ generates arbitrary redefinitions of 
the coordinate $q(t)$, while in this case the first-class Hamiltonian 
$H$ vanishes, $H=0$. An admissible phase space gauge fixing condition 
is $\chi(q,p)=q-q_i$, $q_i$ being some initial value for $q(t)$. At the 
quantum level, and when taking due account of a possible nontrivial U(1) 
holonomy\cite{JG5} for the representation of the Heisenberg algebra 
$[\hat{q},\hat{p}]=i\hbar$, it turns out that the factor $N$ is quantised 
and that the physical spectrum is reduced to a single $\hat{p}$-eigenstate. 
When computing the BRST invariant matrix elements (\ref{eq:ME}) of interest 
for the choice of gauge fixing (\ref{eq:Psi2}) with $\chi=q-q_i$, the 
admissibility of this gauge fixing is confirmed, once again up to a 
normalisation factor stemming from the ghost and Lagrange multiplier sectors 
which is explicitly dependent on the parameters $\rho$ and $\beta$ appearing 
in $\Psi$. In particular, and as is also the case with the result established 
in (\ref{eq:MEadmissible}), the BRST invariant matrix elements (\ref{eq:ME}) 
vanish identically in the limit $\beta\rightarrow 0$,
including those for $k_i=k_{0,i}$ and $k_f=k_{0,f}$ which should
correspond to the nonvanishing physical ones in (\ref{eq:prop1}).

Hence, contrary to the usual statement of the Fradkin-Vilkovisky theorem,
the BRST/gauge invariant BFV path integral is not totally independent of the
choice of gauge fixing ``fermion" $\Psi$. This note revisited once again this
issue, with two main conclusions. First, for a general class of gauge fixing
``fermions", it identified a general criterion for admissibility within a 
simple general class of constrained systems with a single first-class 
constraint which commutes with the first-class Hamiltonian. This criterion 
is in perfect agreement with the conclusions of explicit counter-examples 
to the usual statement of the FV theorem, and it may be seen as a 
continuation of the work in Ref.\cite{FS}. Second, the basic reasons why 
the general argument claiming to establish complete independence of 
the BFV-PI on the gauge fixing ``fermion" is unwarranted in the case of 
the associated BRST invariant matrix elements, have been addressed in simple 
terms. It has been shown that the lack of total independence from $\Psi$ of 
the BFV-PI arises because, whereas the action of the anticommutator 
$\{\hat{\Psi},\hat{Q}_B\}$ on BRST invariant states is unambiguous, that 
of the operators $\hat{\Psi}\hat{Q}_B$ and $\hat{Q}_B\hat{\Psi}$
separately is not.

These conclusions were reached by two separate routes, namely by working
either with the plane wave representation on the real line for the
Lagrange multiplier sector Heisenberg algebra, or else by compactifying
that sector onto a circle in order to avoid having to deal with
non-normalisable states and a continuous spectrum of eigenstates. 
In the first approach, it was shown that
due to the distribution-valued character of the relevant matrix
elements, usual arguments claiming to establish complete independence
from $\Psi$ have to be considered with greater care, thereby confirming
the lack of total independence, even though manifest gauge invariance is
preserved throughout. In the compactified approach, it was shown that the usual
argument is beset by another ambiguity, namely the fact that the Lagrange
multiplier operator $\hat{\lambda}$ maps outside of the domain of
normalisable states for which the conjugate operator $\hat{\pi}$ is
self-adjoint, inducing further crucial surface terms which are ignored by the
usual argument. Incidentally, were it not for such subtle points,
the usual statement of the FV theorem would be correct, so that
the BFV-PI would always be vanishing, irrespective of the choice of gauge
fixing, clearly an undesirable situation since the correct quantum evolution
operator could then not be reproduced. This is explicitly illustrated by
the fact that using the compactification regularisation and
in the limit $\beta\rightarrow 0$, the BFV-PI vanishes for the choices 
(\ref{eq:Psi}) and (\ref{eq:Psi2}), and this independently of
the functions $F(\lambda)$ or the parameter $\rho$. Indeed for these two 
choices, it is precisely the parameter $\beta$ which controls any 
contribution from the gauge fixing ``fermion" to the BFV-PI.

The actual and precise content of the FV theorem is already described in 
the Introduction. As mentioned there, its relevance is really within
a nonperturbative context, while for ordinary perturbation theory,
there is no reason to doubt that the BFV-PI integral should be 
independent of the gauge fixing fermion\cite{Voronov}. However, the
subtle and difficult issues raised by the correct statement of the
Fradkin-Vilkovsky theorem are certainly to play an important role in
the understanding of nonperturbative and topological features of
strongly interacting nonlinear dynamics, such as that of Yang-Mills theories.

Faced with this situation, it thus appears that the admissibility of any given
gauge fixing procedure must be addressed on a case by case basis, once a
dynamical system is considered. In particular, this requires the knowledge
of the modular space of gauge orbits of the system, in general a difficult
problem in itself. However, it should be recalled that any quantisation
procedure of a constrained system not involving any gauge fixing procedure,
such as that based on the physical projector\cite{Klaud1} which is set 
within precisely Dirac's quantisation approach only, avoids having 
to address these difficult problems of identifying modular space and 
assessing admissibility. Indeed, through the physical projector approach, 
an admissible covering of modular space is always achieved 
implicitly\cite{JG4}, as illustrated for example in (\ref{eq:prop2}).

\vspace{20pt}

\noindent{\bf Acknowledgements}

One of the authors (JG) wishes to express his grateful thanks to 
Profs. Hendrik B. Geyer and Frederik G. Scholtz, as well as the Department 
of Physics at the University of Stellenbosch and the Stellenbosch 
Institute for Advanced Study (STIAS), for their most enjoyable and warm 
hospitality while this work was performed. The visit of JG to South Africa 
was supported by a travel grant from the National Scientific Research 
Fund (F.N.R.S.), Belgium. The work of JG is partially supported by the 
Federal Office for Scientific, Technical and Cultural Affairs (Belgium) 
through the Interuniversity Attraction Pole P5/27. FGS acknowledges support 
from the National Research Foundation of South Africa (NRF).

\clearpage

\end{document}